\documentclass[twocolumn]{aastex631}

\shorttitle{Circumbinary disks with an outer companion}
\shortauthors{R. G. Martin et al.}
\begin{document}

\title{Circumbinary disk evolution in the presence of an outer companion star}

\author{Rebecca G. Martin}
\author{Stephen Lepp}
\affil{Nevada Center for Astrophysics, University of Nevada, Las Vegas,
4505 South Maryland Parkway, Las Vegas, NV 89154, USA}
\affil{Department of Physics and Astronomy, University of Nevada, Las Vegas,
4505 South Maryland Parkway, Las Vegas, NV 89154, USA}
\author{Stephen H. Lubow}
\affiliation{Space Telescope Science Institute, 3700 San Martin Drive, Baltimore, MD 21218, USA}
\author{Matthew A. Kenworthy}
\affiliation{Leiden Observatory, University of Leiden, PO Box 9513, 2300 RA Leiden, The Netherlands}
\author{Grant M. Kennedy}
\affiliation{Department of Physics and Centre for Exoplanets and Habitability, University of Warwick, Gibbet Hill Road, Coventry CV4 7AL, UK}
\author{David Vallet}
\affiliation{Department of Mechanical Engineering, University of Nevada, Las Vegas, 4505 South Maryland Parkway, Las Vegas, NV 89154, USA}
%\affil{Department of Physics and Astronomy, University of Nevada, Las Vegas,
%4505 South Maryland Parkway, Las Vegas, NV 89154, USA}

%\author[0000-0002-0786-7307]{Greg J. Schwarz}
%\affil{American Astronomical Society \\
%2000 Florida Ave., NW, Suite 300 \\
%Washington, DC 20009-1231, USA}

%\author{August Muench}
%\affiliation{American Astronomical Society \\
%2000 Florida Ave., NW, Suite 300 \\
%Washington, DC 20009-1231, USA}
%\collaboration{(AAS Journals Data Scientists collaboration)}



\begin{abstract}
We consider a hierarchical triple system consisting of an inner eccentric binary with an outer companion.
A highly misaligned circumbinary disk around the inner binary is subject to two competing effects: (i) nodal precession about the inner binary eccentricity vector that leads to an increase in misalignment (polar alignment) and (ii) Kozai-Lidov (KL) oscillations of eccentricity and inclination driven by the outer companion that leads to a reduction in the misalignment. The outcome depends upon the ratio of the timescales of these effects. If the inner binary torque dominates, then the disk aligns to a polar orientation. If the outer companion torque dominates, then the disk undergoes KL oscillations. In that case, the highly eccentric and misaligned disk is disrupted and accreted by the inner binary, while some mass is transferred to the outer companion. However, when the torques are similar, the outer parts of the circumbinary disk can  undergo large eccentricity oscillations while the inclination remains close to the polar orientation. The range of initial disk inclinations that evolve to a polar orientation is smaller in the presence of the outer companion. Disk breaking is also more likely, at least temporarily, during the polar alignment process. The stellar orbits in HD~98800 have parameters such that polar alignment of the circumbinary disk is expected. In the absence of the gas, solid particles are unstable at much smaller radii than the gas disk inner tidal truncation radius because  KL driven eccentricity leads to close encounters with the binary.
\end{abstract}



\section{Introduction}

%There are now two polar circumbinary gas disks that have been observed, HD~98800 \citep{Kennedy2019} and V773 Tau~B. 

Recent studies have investigated the dynamics of circumbinary disks that are initially
misaligned with respect to the binary orbital plane \citep[e.g.,][]{Nixonetal2013,Facchinietal2013,Aly2015, Martin2017, Lubow2018, Zanazzi2018,  Cuello2019,Smallwood2019, Smallwood2020}.  There are several observed cases of such disks. 
The quadruple star system HD~98800 consists of two binaries that orbit each other with a polar circumbinary gas disk around one of the binaries \citep{Kennedy2019}. In this work, we consider generally the effect of an outer companion (that is a binary in the case of HD~98800) on the dynamics of a highly misaligned circumbinary disk.

In a triple system composed of an inner binary with an outer companion, in order for the stars to be in orbits that are stable against Kozai-Lidov oscillations of inclination and eccentricity \citep{Zeipel1910,Kozai1962,Lidov1962, Naoz2016,Hamers2021}, the orbit of the inner binary must be at an inclination of less than about $40^\circ$ to the outer companion orbital plane.  A polar circumbinary disk is inclined by $90^\circ$ to the inner binary orbit and therefore must be at an inclination greater than about $50^\circ$ to the outer companion orbit. If the inner binary were replaced by a single star, the outer component would cause KL oscillations of such a highly misaligned disk \citep{Martinetal2014,Fu2015,Fu2015b}. During these oscillations, the inclination and eccentricity of the disk are exchanged and the level of disk misalignment is reduced \citep{Martin2016}. The disk is able to respond globally provided that bending waves (which propagate at half the sound speed) can communicate the warp across the radial extent of the disk in less than the disk precession timescale. This occurs for thicker discs where the sound speed is sufficiently high \citep{Papaloizou1995,Larwoodetal1996}.  In order for the disk to remain in a polar configuration, it must be stabilized against KL oscillations. Stabilization is possible through the effects of the inner binary.

The inner binary drives nodal libration of test particle orbits \citep{Farago2010,Doolin2011,Naoz2017,Chen2019}  meaning that a circumbinary disk  undergoes nodal precession about the eccentricity vector of the inner binary. Dissipation within the disk leads to polar alignment where the disk angular momentum is aligned to the binary eccentricity vector \citep{Aly2015, Martin2017,Martin2018,Zanazzi2018,Cuello2019}.   
 
 With both an inner binary and an outer companion, there are now two competing effects on the disk. \cite{Verrier2009} examined this problem for test particle orbits and found that particles close to the inner binary are stabilized against KL oscillations when the nodal libration period is shorter than the KL oscillation timescale. 
 
 In this work, for the first time, we examine the effect of an outer companion star on the evolution of a polar circumbinary disk. In Section~\ref{test} we begin by considering test particle orbits. Particles that undergo KL oscillations are unstable and are ejected through close encounters with the inner binary.
 %We show that if the KL timescale is shorter than the nodal precession timescale that the particles are unstable.  
 In Section~\ref{hydro} we present hydrodynamical gas disk simulations. We find that the gas disk is stable much farther out than the particle orbits but the qualitative behaviour of the disk is similar to a test particle if one of the torques dominates the other.  However, when the inner binary and outer companion have similar magnitude effects, the disk can be both in a polar configuration and undergoing KL eccentricity oscillations. We consider some analytic estimates for the timescales in Section~\ref{analytic} and  draw conclusions in Section~\ref{conc}.

\section{Test particle dynamics}
\label{test}

\begin{figure*}
\begin{center}
\includegraphics[width=1.75\columnwidth]{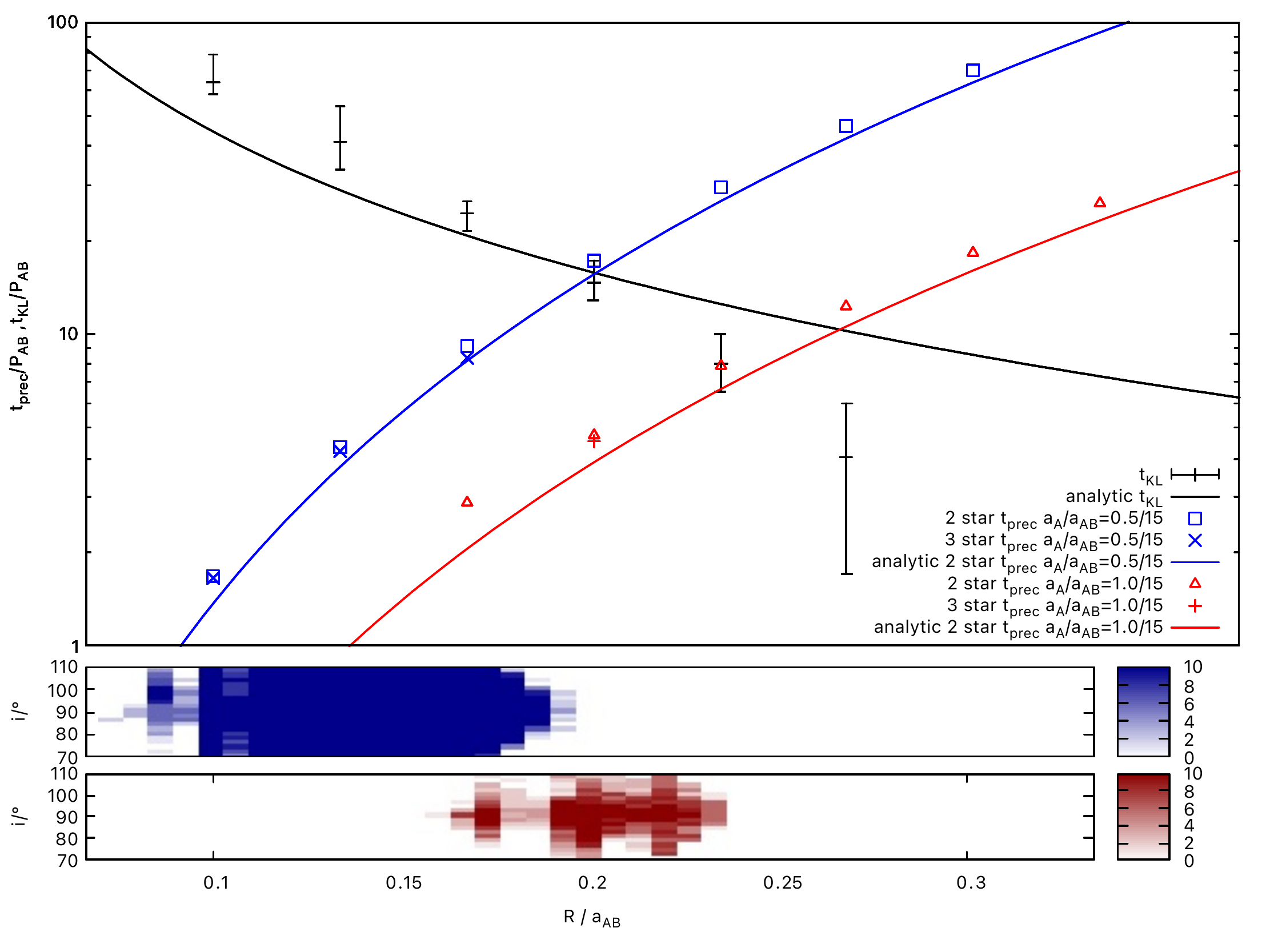}
	\end{center}
    \caption{Test particle orbits. Top panel: Particle libration and KL timescales. 
    %{\bf the vertical axis label should be timescale or $t_{\rm prec}/P_{\rm AB}, t_{\rm KL}/P_{\rm AB}$ as in Fig. 5} 
    The black line shows the KL timescale given by equation~(\ref{KL}). The blue and red lines show the nodal libration periods given by equation~(\ref{prec}) for $a_A/a_{AB}=0.5/15$ and $a_A/a_{AB}=1/15$, respectively.  Each point shows a numerical test particle simulation result. The error bars on the KL periods show the range of values with ten different starting true anomaly  values. The size of the error bars on the libration periods would be smaller or equal to the size of the points.} The lower panels show test particle stability for a range of initial inclinations. Each pixel consists of 10 simulations that begin with different values of true anomaly.  
    \label{testp}
\end{figure*}

We examine test particle orbits with analytic approximations and the $n$-body code {\sc rebound} \citep{Rein2012} using the WHFast integrator.  The inner binary is equal mass $M_{Aa}=M_{Ab}=0.5\,M_A$, where the total mass is $M_A=M_{Aa}+M_{Ab}$. The inner binary orbit has semi-major axis $a_A$ and eccentricity $e_A$. The outer companion is treated a single star with mass $M_B=M_A$ in a circular orbit with the inner binary with semi-major axis $a_{AB}$.  

\subsection{Inner binary}

First, we examine test particle orbits around the inner binary without the outer companion.  Tilt oscillations of a  test particle on a nearly polar orbit at a distance $R$ from the inner binary, as a result of the nodal libration, occur with a period given by
\begin{equation}
    t_{\rm prec}=\frac{2 \pi}{\omega_{\rm prec}},
\end{equation}
where
\begin{equation}
    \omega_{\rm prec} = k \frac{M_{Aa}M_{Ab}}{M_{A}^2} \left(\frac{a_{A}}{R}\right)^{7/2} \Omega_{A},
    \label{prec}
\end{equation}
\begin{equation}
k=\frac{3 \sqrt{5}}{4}e_{A} \sqrt{1+4e_{A}^2}
\end{equation}
\citep{Farago2010, Lubow2018}, the angular frequency of the inner binary is $\Omega_{A}=2 \pi /P_{A}$, and the orbital period of the inner binary is $P_{A}$. The relation between the nodal libration period and distance is shown in the blue and red lines in the upper panel of Fig.~\ref{testp} for $a_A/a_{AB}=0.5/15$ and $a_A/a_{AB}=1.0/15$, respectively, both with $e_A=0.5$. The separation is scaled by the separation of the outer companion, $a_{AB}$, and time by the orbital period of the outer companion, $P_{AB}$. Even though the outer companion does not affect $t_{\rm prec}$, we do this for easier comparison later.  The open square and open triangular points  show the nodal libration period for test particle orbits around the inner binary that are initially misaligned by $70^\circ$.   There is good agreement between the numerical simulations and the analytic approximation. The agreement gets better for larger $R/a_{AB}$ as expected, since the quadrupole
approximation used in deriving equation (\ref{prec}) becomes more accurate. The agreement also gets better at higher initial inclination where the orbit is closer to polar.

\subsection{Outer companion}

With an outer companion, the timescale for KL oscillations of the test particle (replacing the inner binary by a single star of mass $M_A$) is given approximately by
\begin{equation}
    t_{\rm KL}=\frac{M_A+M_B}{M_B}\frac{P_{AB}^2}{P_{\rm p}} (1-e_{AB}^2)
    \label{KL}
\end{equation}
\citep{Kiseleva1998,Ford2000}, where  $P_{\rm p}=2\pi/\sqrt{GM_A/R^3}$ is the orbital period of the test particle around the A star and $e_{AB}$ is the eccentricity of the outer companion orbit. We note that the equation (\ref{KL}) is derived under the quadrupole approximation in which the KL oscillation period for particles on initially circular orbits is formally infinite due to a logarithmic divergence \citep[e.g.,][]{Lubow2021}. 
This equation gives an estimate for the period to within factors of order unity.  
The period $t_{\rm KL}$ is plotted as a function of $R/a_{\rm AB}$ in the  black line in the upper panel of Fig.~\ref{testp}. The black points  show the KL oscillation periods obtained from test particle simulations with initial inclination of $70^\circ$ in which the A binary is replaced by a single star. %and the B companion.
The KL period is very sensitive to the initial conditions and so we average the periods over ten different values of the initial true anomaly.  The higher the initial inclination, the better the agreement with equation~(\ref{KL}).  For test particles that are at orbital radii $R/a_{AB}\gtrsim 0.2$, the analytic KL timescale is higher than the numerical value of the KL period.  The analytic timescale is derived in the limit that the semi-major axis of the particle is much smaller than the binary semi-major axis.

\subsection{Inner binary and outer companion}

We now consider test particle simulations that include both the inner binary and the outer companion that orbit in the same orbital plane.
%The inner binary has $e_A=0.5$ and the outer companion is in a circular orbit. 
The red crosses and blue Xs in the upper panel of Fig.~\ref{testp} show the nodal libration periods for simulations that include the three stars. The libration period is not significantly altered by the outer star. However, the range of initial semi-major axes of the particle for which stable orbits exist is greatly reduced, as seen by the limited range that the crosses and the Xs appear in the plot.

The lower two panels show stability maps for test particles in the three star systems. The upper map is for $a_A/a_{AB}=0.5/15$ and the lower map is for $a_A/a_{AB}=1/15$. Each pixel is colored according to number of stable cases among 10 simulations that have different (equally spaced) initial true anomaly  values.  We run each simulation for a time of $50,000\,P_{A}$ and define a particle to be unstable if its eccentricity becomes larger than 1, or its semi-major axis becomes large ($a_{\rm p}>10\,a_{A}$) or small ($a_{\rm p}<a_{A}$) \citep[see for example][]{Quarles2018,Chen2020}.

The stable region for test particles extends out to the radius where the numerically determined KL period becomes smaller than the nodal precession period. This was first seen by \cite{Verrier2009} with regards to the HD~98800 system.  Note that for the wider binary ($a_A/a_{AB}=1/15$, lower panel), the outer edge is close to where the numerical KL period is equal to the nodal libration period while for the closer inner binary, the analytic KL timescale is a good approximation.

%Critical inclination between circulating and librating.

\section{Hydrodynamical gas disk dynamics}
\label{hydro}

\begin{figure*}
\begin{center}
\includegraphics[width=0.9\columnwidth]{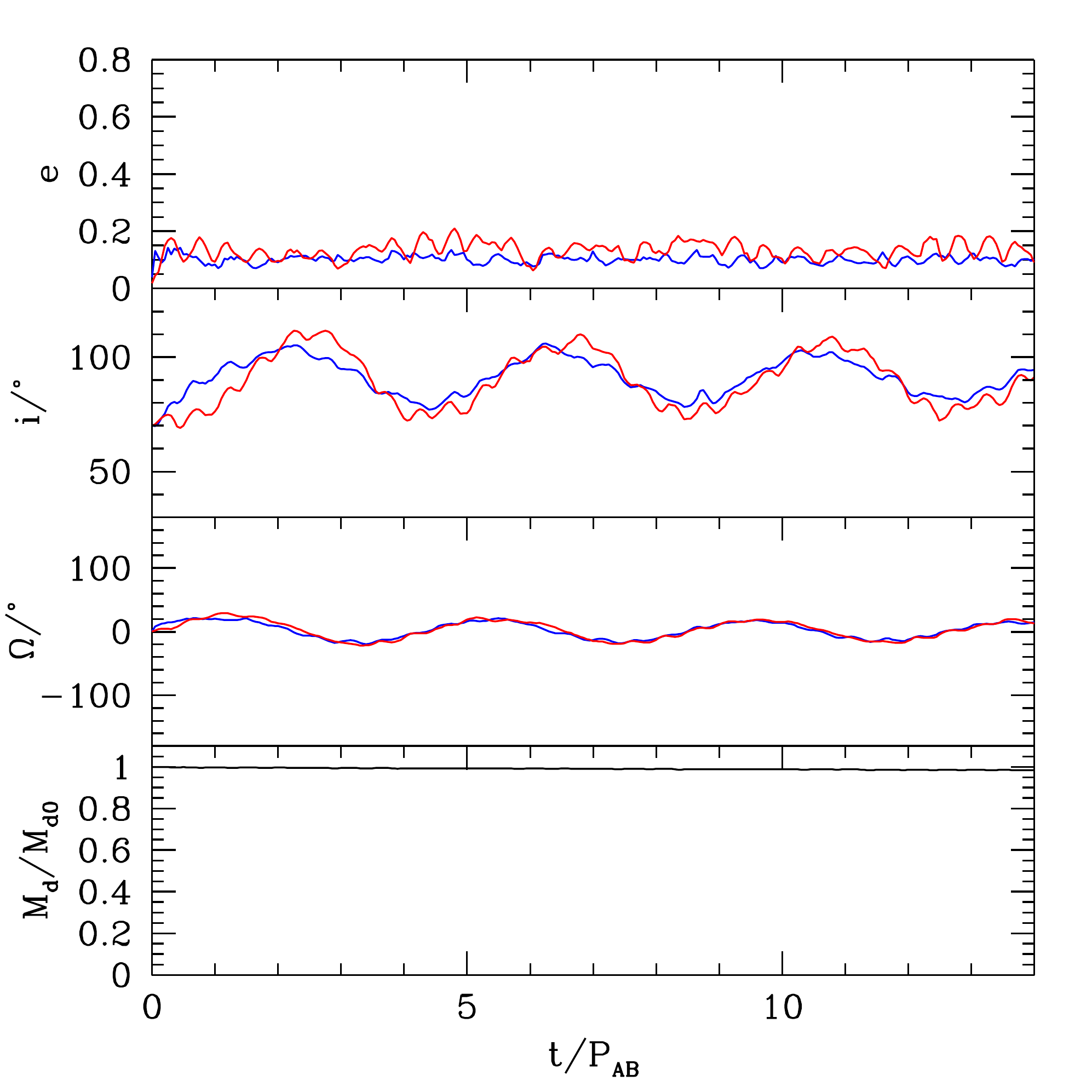}
\includegraphics[width=0.9\columnwidth]{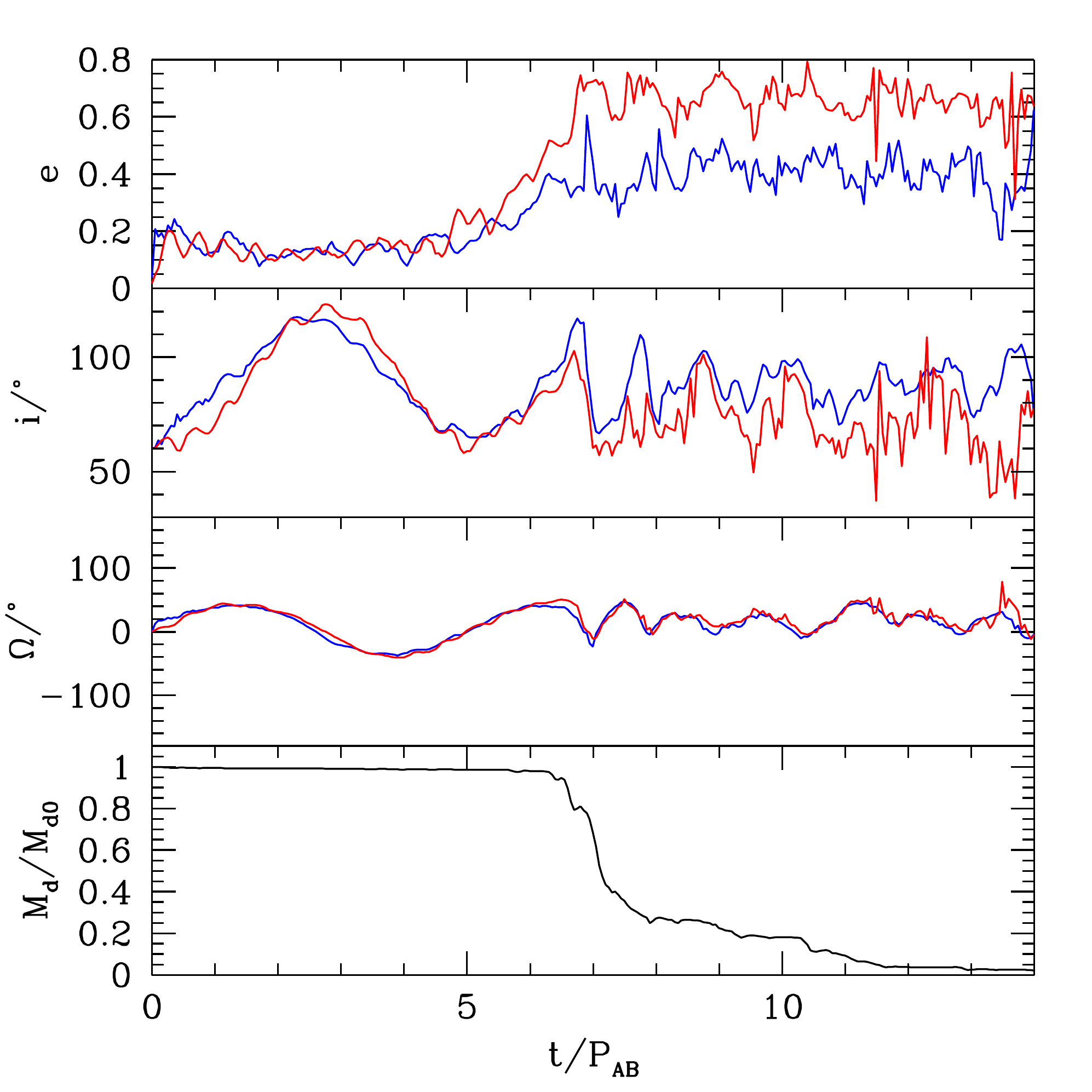}
	\end{center}
    \caption{Hydrodynamical gas disk simulations for the wide inner binary with $a_{A}/a_{AB}=1/15$ and $e_A=0.5$ with initial disk inclination of $70^\circ$ (left) and $60^\circ$ (right).  The blue lines show the inner disk at $R=2\, a_{A}$ and the red lines show the outer disk at $R=(1/3)\,a_{AB}$.  The time is in units of the orbital period of the outer companion. From top to bottom, the panels show the disk eccentricity, inclination, longitude of ascending node and the circumbinary disk mass.}  
    \label{angles}
\end{figure*}

We use the smoothed particle hydrodynamics (SPH) code {\sc phantom} \citep{Price2010,Price2018} to model the evolution of a hydrodynamical gas disk with an inner binary and an outer companion. Misaligned disks in binaries have been widely studied with this code \citep[e.g.][]{Nixonetal2013,Nealon2018}. We first consider the same triple star parameters as in Section~\ref{test}.
The accretion radii of the inner binary stars are $0.25\,a_{A}$ while the accretion radius of the outer companion star is $(0.25/15)\,a_{AB}$. Particles that move inside of this radius are accreted and their mass and angular momentum are added to the sink particle \citep{Bateetal1995}. 

The disk has a mass of $0.001\,M_A$ with $500,000$ particles initially. Since the disk mass is low we ignore self-gravity of the disk. The particles in the gas disk are initially distributed with a power law surface density $\Sigma \propto R^{-3/2}$ between $R_{\rm in}=2\, a_A$ and $R_{\rm out}=(1/3)\, a_{AB}$. The inner disk edge is chosen to be close to the tidal truncation radius from the inner binary. This is smaller around a polar binary compared to a coplanar binary \citep{Lubow2018,Franchini2019inner}. The outer disk edge is chosen to be close to the  tidal truncation radius from the outer companion \citep{Artymowicz1994}. Note that an inclined disk has a larger tidal truncation radius than a coplanar disk \citep{Lubowetal2015,Miranda2015}.  The disk is locally isothermal with sound speed $c_{\rm s}\propto R^{-3/4}$ and we take $H/R=0.1$ at the initial disk inner radius. This is chosen in order to make the shell averaged smoothing length per scale height, $\left<h\right>/H$, constant with radius \citep{Lodato2007}. The \cite{SS1973} viscosity $\alpha$ parameter is 0.01 and is implemented by adapting the SPH artificial viscosity with $\alpha_{\rm AV}=0.34$ (with $a_A/a_{AB}=0.5/15$) or $\alpha_{\rm AV}=0.44$ (with $a_A/a_{AB}=1/15$) and $\beta_{\rm AV}=2$ \citep{Lodato2010}. The disk is resolved with  $\left<h\right>/H=0.29$
%{\bf Should this be $\left<h\right>/H=0.29 $ ? Yes, thank you!} 
(with $a_A/a_{AB}=0.5/15$) or $\left<h\right>/H=0.23$ (with $a_A/a_{AB}=1/15$). Since $H/R \gg \alpha$, disk warping occurs in the bending wave regime \citep{Papaloizou1983}. We bin the particles into 100 bins in spherical radius and average the properties of the particles in each bin. The inclination and the longitude of ascending node are calculated in a frame relative to the inner binary \citep[see equations 1 and 3 in][]{Chen2019}. 

\subsection{Wide inner binary}
\label{widebinary}
We first consider simulations with a relatively large ratio of the binary semi-major axes, $a_A/a_{AB}$, so that the disk is radially narrow. The left panel of Fig.~\ref{angles} shows the evolution of a disk around a binary with $a_A/a_{AB}=1/15$ and $e_A=0.5$ that is initially at an inclination of $70^\circ$ to the binary orbit. The disk undergoes tilt oscillations and aligns towards a polar inclination. The behavior at the two radii in the disk that are plotted is similar which indicates that there is little warping in the disk. This is a natural consequence of being in the bending wave regime. The communication timescale, $t_{\rm c} \sim 2R/c_{\rm s} \sim 2(H/R)^{-1}/\Omega_{\rm out}$, where $\Omega_{\rm out}$ is the Keplerian frequency in the outer parts of the disk,  is roughly  $0.9\, P_{\rm AB}$. Since the tilt oscillations seen in the left panel of Fig.~\ref{angles}  occur over longer timescales, the disk behaves rigidly with little warping. There is also little disk eccentricity growth in this simulation, since the inner binary torque is dominating the outer companion torque and suppressing the KL oscillations. The parameters of the inner binary and disk are exactly the same as those in the simulation presented in \cite{Martin2017} except that the initial disk inclination here is $70^\circ$, compared to $60^\circ$ in \cite{Martin2017}. The behavior is qualitatively very similar except that the tilt oscillations occur on a shorter timescale.

The right hand panel of Fig.~\ref{angles}  shows a simulation with the same parameters as the left panel but a lower inclination of $60^\circ$. At a lower inclination, the nodal libration period is longer. The parameters are exactly those presented in \cite{Martin2017} in which the disk aligned to polar without the outer companion star. Initially the disk undergoes tilt oscillations with a reduced  period in the presence of the outer companion. The period of the first oscillation is about $4.75\,P_{AB}\approx 200\,P_A$ with the outer companion, while it was about $250\,P_{A}$ without the outer companion \citep[see Fig.~1 in][]{Martin2017}.  The test particle nodal libration timescale is not significantly affected by the companion star (see Section~\ref{test}) and so this is likely because the disk is able to spread out to larger radii without the outer companion. The libration period for a rigid disk is set by an appropriate average of the precession periods over the disk radial extent. However, after just one libration period, the KL effect from the companion dominates the dynamics and the disk becomes highly eccentric. 
The eccentricity growth is uniform over the disk radial extent. This leads to a strong interaction with the inner binary that largely disrupts the disk. Some of the disk material is transferred to the outer companion \citep{Franchini2019} and it ends up with a more massive disk than the inner binary. 
%{\bf This is true even though the inner binary accretes more mass (next sentence)?}
%{\bf how much goes to the outer companion vrs. in the inner binary?} 
At a time of $14\,P_{AB}$, the inner binary has accreted 73\% of the initial disk mass and the remaining circumbinary disk contains about 2\%. The companion has accreted about 6\% and the  disk left around it contains about 10\%. The remaining 9\% of the material has been flung out to large radius or is forming  a circumtriple disk. The range of initial inclinations that lead to polar alignment is smaller in the presence of an exterior binary companion.

\subsection{Close inner binary}

\begin{figure}
\begin{center}
\includegraphics[width=0.9\columnwidth]{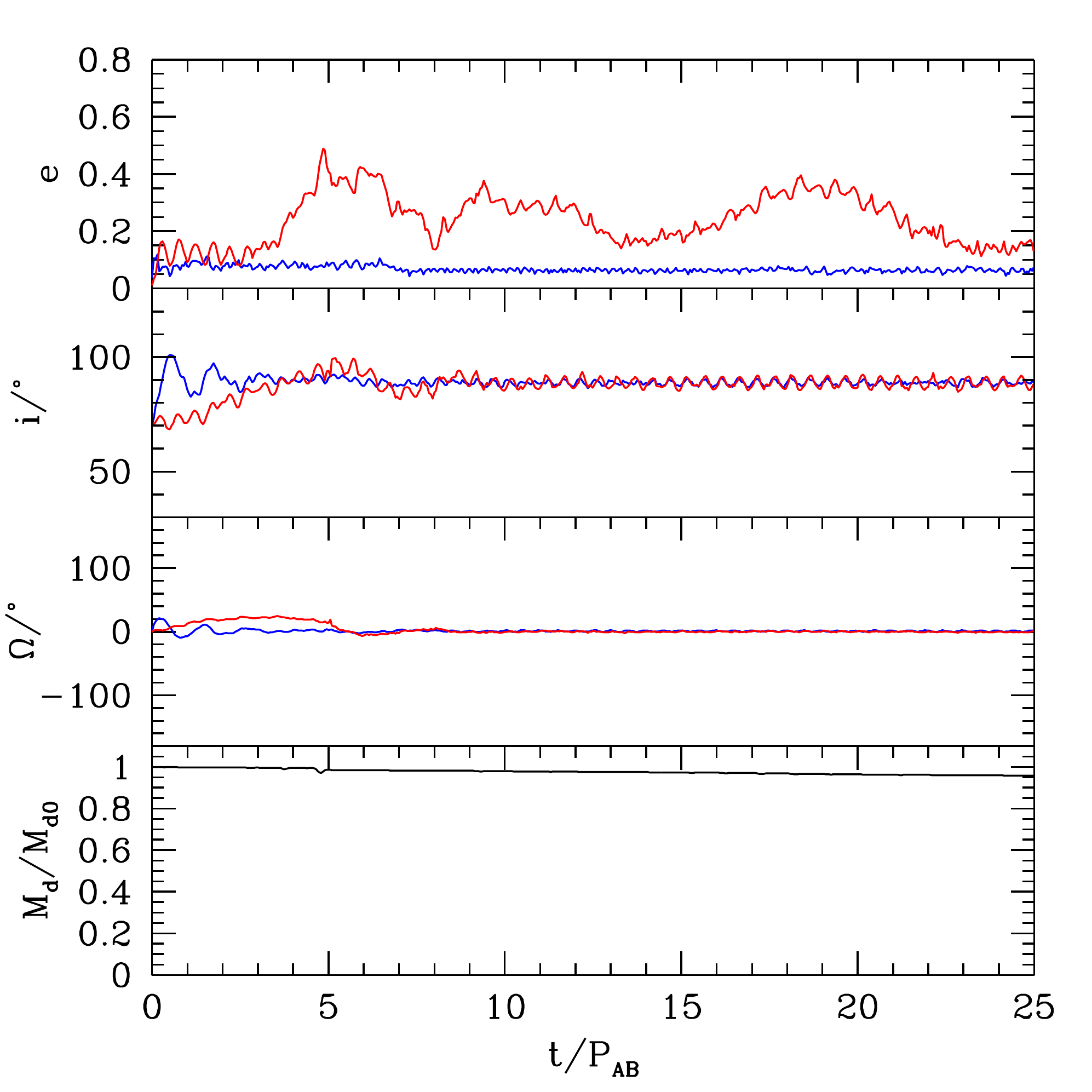} 
	\end{center}
    \caption{Same as the left panel of Fig.~\ref{angles} except for  $a_{A}/a_{AB}=0.5/15$.}  
    \label{angles2}
\end{figure}

\begin{figure} 
\begin{center}
\includegraphics[width=\columnwidth]{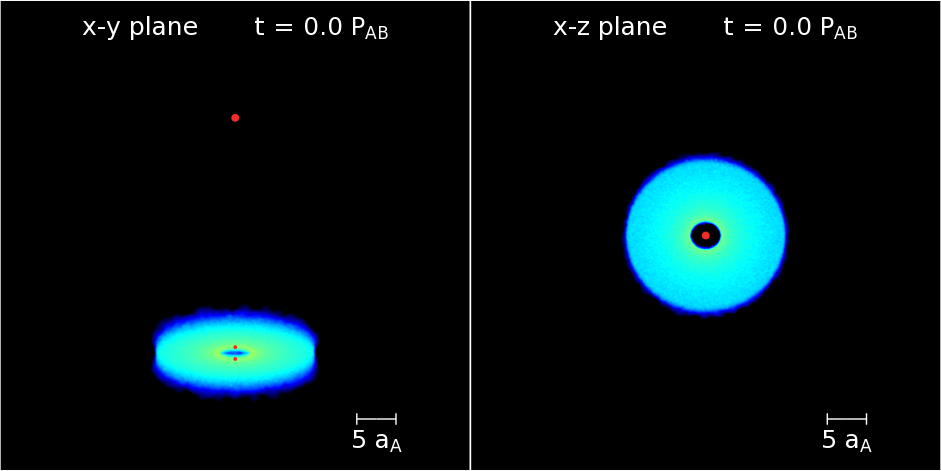}
\includegraphics[width=\columnwidth]{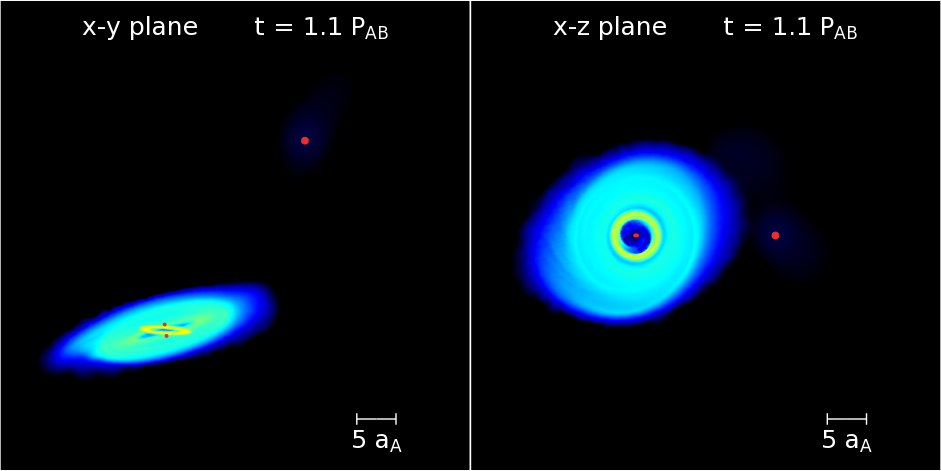}
\includegraphics[width=\columnwidth]{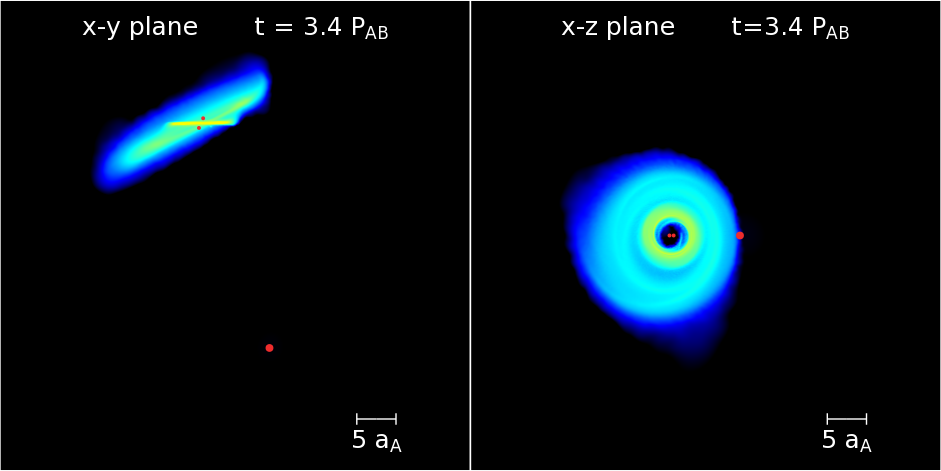}
\includegraphics[width=\columnwidth]{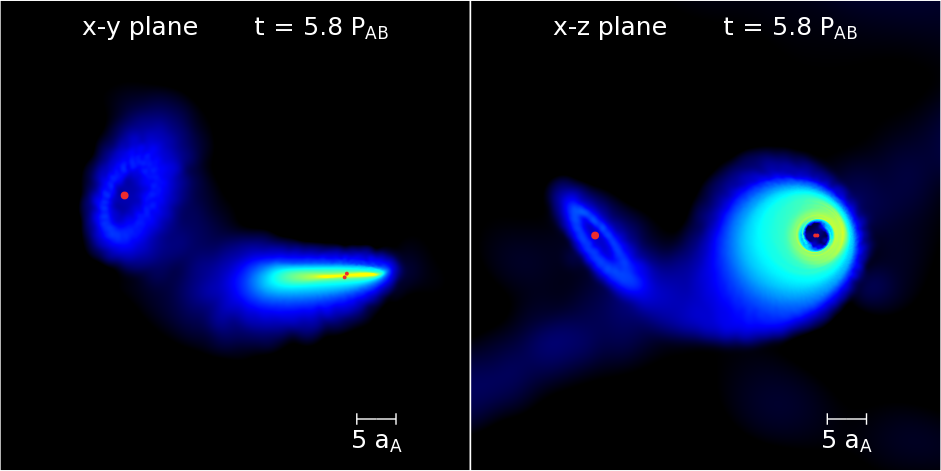}
	\end{center}
    \caption{Close inner binary simulation with $a_A/a_{AB}=0.5/15$ and $e_A=0.5$ at times of $t=0$, $1.1$, $3.4$ and $6\,P_{AB}$ from top to bottom. The red circles show the three stars scaled to the size of their sink radius. The left panels show the $x-y$ plane in which both the inner binary and the outer companion orbit. The right panels show the $x-z$ plane. In the final time, the circumbinary disk is perpendicular to the inner binary and the outer companion orbits.  The highly eccentric outer parts of the circumbinary disk have transferred material to the outer companion and it has a low mass disk.
    %331 time of 673PA
    }
    \label{splash}
\end{figure}

 Fig.~\ref{angles2} shows a simulation with a smaller inner binary separation of $a_A/a_{AB}=0.5/15$ and $e_A=0.5$ with initial disk inclination of $70^\circ$. The initial disc configuration is shown in the top panels of Fig.~\ref{splash}. The tilt oscillation period is shorter for this closer binary and the inner parts of the disk undergo rapid polar alignment. The disk is strongly warped by a time of about $1\,P_{AB}$ where there is about a $20^\circ$ difference over a radial range of $0.03\,\rm a_{AB}$ in both the inclination and the nodal phase angles (see the second row of Fig.~\ref{splash}). The warp strength then weakens and the warp location propagates outwards in time (see the third row of Fig.~\ref{splash}). Note that with higher resolution simulations, the disk may look more cleanly broken rather than warped \citep{Nealon2015}. We see slower polar alignment in the outer parts of the disk compared to the inner disk.  However, the outer parts of the disk undergo significant eccentricity growth as a result of the KL effect. The eccentricity growth is not uniform with radius but increases with separation from the inner binary. These eccentricity oscillations of the polar disk may be long lived since the disk remains in a polar configuration, that is, above the critical inclination required for KL disk oscillations \citep{Lubow2017,Zanazzi2017}. Long lived KL disk oscillations have been seen before when there is a source of high inclination material feeding the formation of circumstellar disks \citep{Smallwood2021}. The lower panels of Fig.~\ref{splash} shows the disk at a time of $t=5.8\,P_{AB}$, near the peak disk eccentricity. The circumbinary disk is polar to both the inner binary and the outer companion orbit while being quite eccentric. A low mass disk can be seen around the companion that has formed as a result of mass transfer during the high eccentricity disk phase \citep[e.g.][]{Franchini2019}.

\section{Global disk timescales}
\label{analytic}

We now estimate analytically the global disk libration period (in the absence of the outer companion)  and the KL timescale (in which we replace the inner binary by a single star). The global precession rate is determined by prcessional torque on the disk divided by its angular momentum, assuming that the disk does not break and behaves rigidly \citep{Papaloizou1995, Larwood1997,Lubow2001}. This is expected in the bending wave regime when the communication timescale is shorter than the global precession timescale. We use equation~(16) in \cite{Lubow2018} to calculate the global disk precession period due to the inner binary and equation~(4) in \cite{Martinetal2014} to calculate the global disk KL timescale due to the outer companion. Note that there is no inclination dependence in these estimates and they are valid close to polar. 
We assume a power law surface density profile distributed between $R_{\rm in}=2.3\,a_A$ and $R_{\rm out}=(5.5/15)\,\rm a_{AB}$. While the inner edge of a polar circumbinary disk is smaller than this value of $R_{\rm in}$ \citep{Franchini2019inner}, the surface density profile tapers close to the binary. We choose this value since it is close to the peak in the surface density profile in our simulations.

\begin{figure}
\begin{center}
\includegraphics[width=0.85\columnwidth]{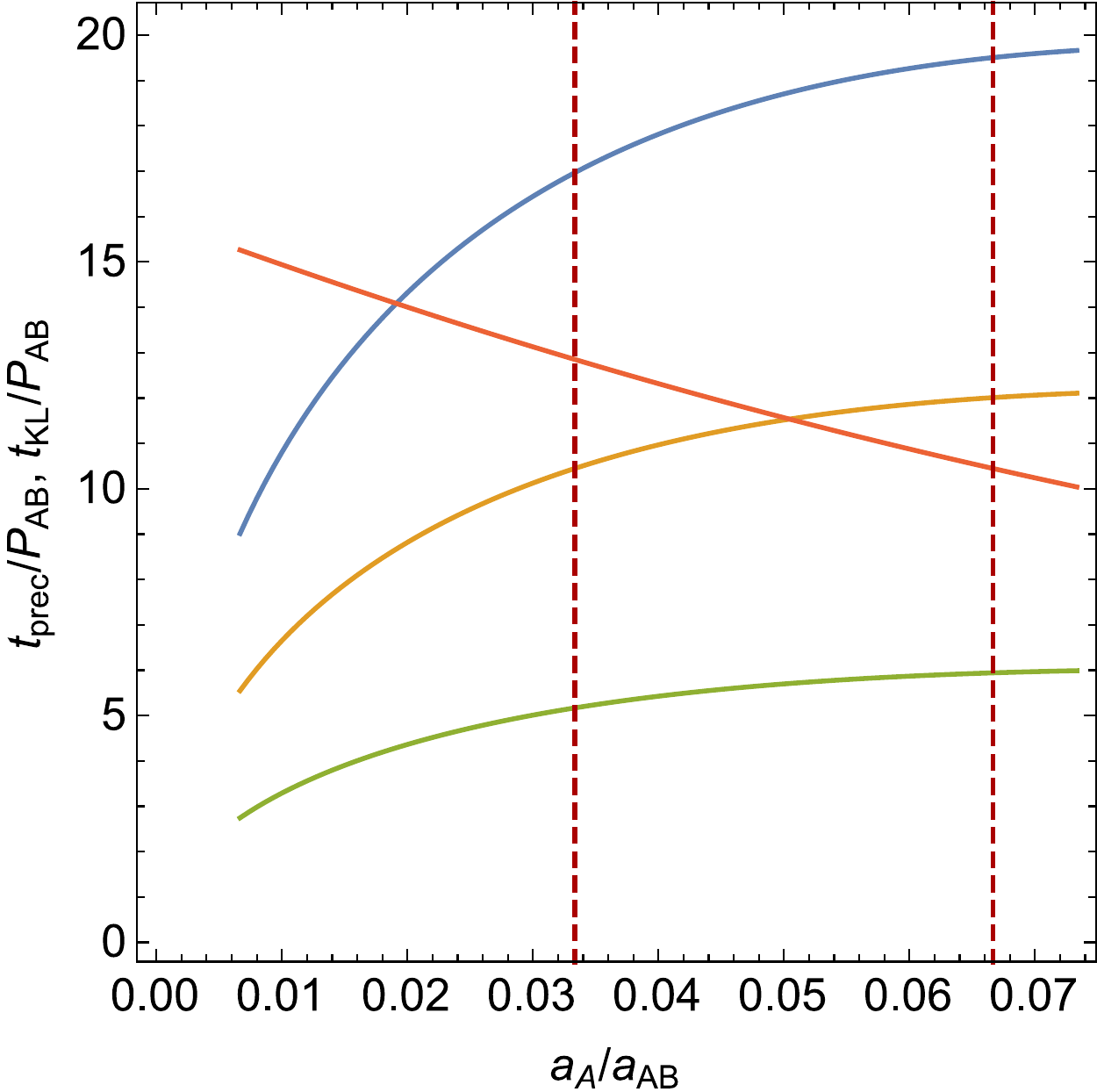}
	\end{center}
    \caption{Analytic estimates for the global precession period around a binary with eccentricity $e_A=0.2$ (blue line), $0.3$ (yellow line) and 0.5 (green line) and the KL oscillation timescale (red line) for a disk with $\Sigma \propto R^{-3/2}$ from $R_{\rm in}=2.3\,a_A$ out to $(1/3)\, a_{AB}$. The vertical dashed lines show the values of the semi-major axis ratios used in the SPH simulations ("close inner binary" on the left and "wide inner binary" on the right).
    }
    \label{fig:analytic}
\end{figure}

Fig.~\ref{fig:analytic} shows these timescales as a function of the ratio of the binary semi-major axes. In our wide binary simulations with $a_A/ a_{AB}=1/15$, the libration period is about $4.75\,P_{AB}$, in rough agreement with this analytic estimate for $e_A=0.5$ (green line). The KL timescale decreases with increasing inner binary separation only because the disk inner radius becomes larger. These analytic timescales provide an estimate of the outcome of a disk simulation if the disk is in good radial communication (when the communication timescale is shorter than the global precession timescale). If the global disk KL timescale is shorter than the global disk nodal libration period then the disk does not remain polar. It undergoes KL oscillations and may be accreted on to the inner binary and transferred to the companion. However, if the nodal libration period is shorter than the KL timescale then we expect the disk to move to polar alignment. 

We also considered some simulations with lower binary eccentricity. For the wide inner binary  ($a_A/a_{AB}=1/15$), with $e_A=0.2$ and 0.3, the disk is destroyed through KL oscillations. This is in agreement with the analytic prediction in Fig.~\ref{fig:analytic}. For the close binary simulation ($a_A/a_{AB}=0.5/15$), we found for $e_A=0.2$ and 0.3 that the disk breaks and much of the material ends up in a polar configuration.  When the disk breaks, the inner ring quickly aligns to polar, while the outer ring undergoes KL oscillations and is disrupted when the disk becomes highly eccentric. The outer ring is accreted on to the stars and forms a disk around the third star. The inner polar ring then spreads outwards and the long term behaviour is similar to that shown in Fig.~\ref{angles2}.   Therefore, disk breaking in a radially wide disk can affect the outcome and help to stabilize the disk against KL oscillations.

%We have only considered systems in which the stars all orbit in the same plane. However, the outer companion could be inclined with respect to the inner binary. This leads to a slightly longer KL timescale for the disk. 

\section{Conclusions}
\label{conc}

A polar circumbinary disk that has an outer companion is subject to two competing dynamical effects. The inner binary causes nodal libration and polar alignment. The outer companion drives KL oscillations that lead to coplanar alignment.   The outcome depends strongly upon the inner binary eccentricity, the initial disk inclination and the ratio of the binary semi-major axes. If the global nodal libration period is shorter than the global KL timescale, the disk moves towards polar alignment. If the KL timescale is shorter than the nodal libration period, the disk becomes highly eccentric and is largely destroyed, either being accreted on to the central binary or forming a disk around the companion. If the timescales are similar, an outer companion to a polar circumbinary gas disk can cause eccentricity growth of the disk while the disk inclination remains in a polar configuration. 
The range of initial inclinations that lead to a polar aligned disk is reduced by the outer companion and disk breaking is more likely, at least temporarily.

The outermost stable particle orbit is significantly closer in than the outer edge of a gas disk. Thus, a polar gas disk can extend to much larger radius than a particle disk.  Solid bodies may be on stable polar orbits while the gas disk is present, but once the gas has dissipated, they may be unstable to KL oscillations and become ejected from the system. 

For the HD~98800 system, the ratio of semi-major axes is $a_{A}/a_{AB}=0.02$ and the eccentricity of the inner binary is $0.79$.  Polar alignment is then expected (see Fig.~\ref{fig:analytic}).  The eccentricity of the outer companion orbit is also relatively large at around 0.52. The high eccentricity of the outer companion leads to a smaller outer truncation radius for the circumbinary disk \citep{Artymowicz1994}. For a power law disk density distribution in radius, the effective radius for the disk precession scales approximately with the periastron separation, 
%$\left<R_{\rm disk}\right>
$R_{\rm eff} \propto (1-e_{AB})$. For fixed stellar masses and outer companion orbital period, the disk KL timescale varies with $(1-e_{AB}^2)^{3/2}/R_{\rm eff}^{3/2} \propto (1+e_{AB})^{3/2}$ (see equation~\ref{KL}). Therefore a higher outer companion eccentricity leads to a longer disk KL oscillation timescale and polar alignment is more likely.

\begin{acknowledgements}

We thank the referee, Daniel Price, for providing useful comments that improved the manuscript. Computer support was provided by UNLV’s National Supercomputing Center.  RGM and SHL acknowledge support from NASA through grants 80NSSC21K0395 and 80NSSC19K0443.
We acknowledge the use of SPLASH \citep{Price2007} for the rendering Fig.~\ref{splash}.

\end{acknowledgements}




\bibliographystyle{aasjournal}
\bibliography{mainapjl} % if your bibtex file is called example.bib



\end{document}